\documentstyle[prl,twocolumn,epsf,rotate,aps]{revtex}
\begin{document}
\draft
\twocolumn[\hsize\textwidth\columnwidth\hsize\csname @twocolumnfalse\endcsname
\title{Spin-triplet superconductivity in Sr$_2$RuO$_4$ probed by Andreev reflection}
\author{F. Laube$^1$, G. Goll$^1$, H. v. L\"ohneysen$^1$, M. Fogelstr\"om$^2$ and 
F. Lichtenberg$^3$}
\address{$^1$ Physikalisches Institut, Universit\"at Karlsruhe,
D-76128 Karlsruhe, Germany\\
$^2$ Institut f\"ur Theoretische Festk\"orperphysik, Universit\"at Karlsruhe,
D-76128 Karlsruhe, Germany\\
$^3$Experimentalphysik VI, Elektronische Korrelationen und Magnetismus,
    Institut f\"ur Physik, Universit\"at Augsburg,\\
D-86135 Augsburg, Germany}
\date{July 08, 1999}
\maketitle
\begin{abstract}
The superconducting gap function of Sr$_2$RuO$_4$ was investigated by means of 
quasiparticle reflection and transmission at the normal conductor-superconductor interface of
Sr$_2$RuO$_4$-Pt point contacts. We found two distinctly different types of $dV/dI$ vs
$V$ spectra either with a double-minimum structure or with a zero-bias conductance anomaly.
Both types of spectra are expected in the limit of high and low transparency, respectively,
of the interface barrier between a normal metal and a spin-triplet superconductor.
Together with the temperature dependence of the spectra this result strongly supports a 
spin-triplet superconducting order parameter for Sr$_2$RuO$_4$.
\end{abstract}
\pacs{74.70.Dd, 71.20.Lp, 73.40.Jn, 74.80.Fp}
]
A superconductor is denoted as unconventional if below the 
transition temperature $T_c$ additional symmetries, e.\,g. the time-reversal symmetry,
are broken besides the gauge
symmetry. Unconventional superconductivity (SC) is reflected by internal degrees of 
freedom of the order parameter (OP) which is determined by the probability
amplitude for a Cooper pair at a given temperature $T$, and a 
non-phononic attractive pair interaction. 
In superfluid $^3$He a spin-triplet pairing state is realized with parallel spin 
$S=1$ and relative orbital momentum $l=1$ of the Cooper pairs and the pair 
interaction is mediated by spin fluctuations \cite{vol90}.
A number of heavy-fermion systems also present evidence for unconventional
SC \cite{hef96}. In the high-$T_c$ cuprates, the OP is 
predominantly singlet $d$ wave $(l=2)$ with some $s$-wave admixture 
\cite{har95,kou97,wei98}.
The layered superconductor Sr$_2$RuO$_4$ with $T_c$ up to 1.5\,K \cite{mae94}
is a prime candidate for $p$-wave SC in an electronic
system as supported by remarkable properties of both the normal and 
superconducting states: (i) The related compounds SrRuO$_3$ and 
Sr$_3$Ru$_2$O$_7$ are both itinerant ferromagnets and spin fluctuations of 
predominantly ferromagnetic character were inferred from $^{17}$O NMR measurements
in Sr$_2$RuO$_4$ \cite{ima98}. However, antiferromagnetic fluctuations were observed by
inelastic neutron scattering \cite{sid99}. (ii) Below $T\approx 25$\,K, 
Sr$_2$RuO$_4$ reflects Fermi-liquid 
behavior in the thermodynamic and transport properties: the 
resistivity exhibits a 
$T^2$ dependence and both linear specific heat and Pauli spin susceptibility
are enhanced by a factor 3-4 with respect to the free-electron model \cite{mae96}.
(iii) The superconducting state is extremely sensitive to disorder and $T_c$
is strongly suppressed by nonmagnetic as well as magnetic impurities \cite{mac98}.
(iv) Furthermore, the specific-heat data indicate a  
residual density of states in the superconducting state at least in  
some samples \cite{mae97}. These hints at unconventional SC suggesting that 
Sr$_2$RuO$_4$ is an 
electronic analogue of $^3$He \cite{ric95}. Indeed, a spin-triplet state has
been identified by Knight-shift measurements \cite{ish98}, and the occurrence
of a spontaneous magnetic field below $T_c$ as revealed by $\mu$SR measurements
\cite{luk98} provides direct evidence for time-reversal symmetry broken 
SC. However, direct experiments to probe the anisotropic
gap structure associated with $p$-wave SC have not yet been performed.
Here we report on the investigation of the gap function of Sr$_2$RuO$_4$
by means of point-contact (PC) spectroscopy well below $T_c$. For current injection
predominantly parallel to the $ab$-plane we observe two distinctly different 
types of $dV/dI$ vs $V$ spectra: (i) a typical double-minimum structure seen
in many superconductors and (ii) a single-minimum structure centered at $V=0$,
i.\,e. a zero-bias conductance peak. The zero-bias anomaly (ZBA) is probably
caused by an Andreev bound state which is a signature 
of unconventional SC with a sign change of 
the pair potential as a function of ${\bbox k_{\rm F}}$. The different structures in $dV/dI$ vs $V$ 
might reflect the transition from
a metallic to a tunneling PC for a $p$-wave superconductor.
We note that Kondo-type
scattering by magnetic impurities in the surface barrier can lead to a
ZBA as well \cite{wol85}. However,
the fact that the observed ZBA vanishes at $T_c$ renders this
possibility unlikely.

The Sr$_2$RuO$_4$ single crystals were grown in air by a modified floating-zone
melting process \cite{lic92}. The superconducting transition temperature 
$T_c=1.02$\,K and the transition width $\Delta T_c^{10-90\%}=35$\,mK were 
determined from bulk resistivity measurements. No ferromagnetic impurity phases 
have been detected by X-ray powder diffraction and magnetization measurements
with a SQUID magnetometer within the resolution of both methods.
Heterocontacts between 
the superconducting sample (S) and a normal-metal (N) counterelectrode Pt 
were established with preferred direction of current injection parallel 
to the $ab$-plane of tetragonal Sr$_2$RuO$_4$. The setup was mounted
inside the mixing chamber of a 
$^3$He/$^4$He dilution refrigerator. Mechanical feedthroughs allowed establishing
and changing of Sr$_2$RuO$_4$-Pt PCs at low T. The 
spectra, i.\,e. the differential resistance $dV/dI$ as a function of applied 
bias $V$, were recorded by standard lock-in technique.

\begin{figure}[b]
\centerline{\epsfysize=0.55\textwidth \epsfbox{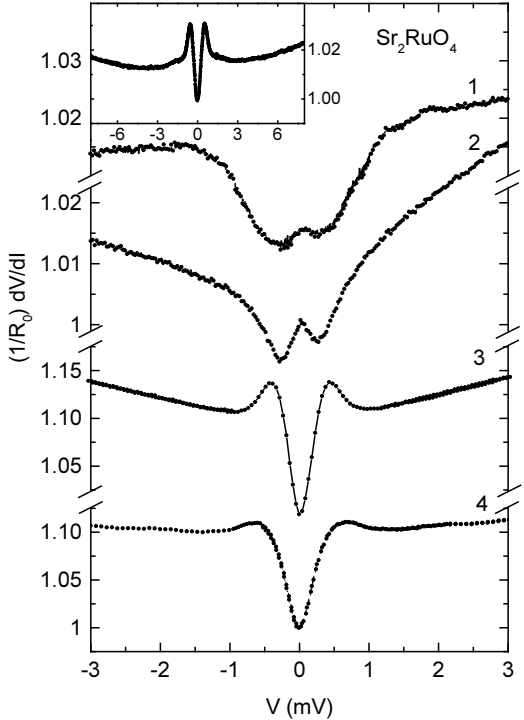}} 
\caption{Differential resistance $dV/dI$ vs bias voltage $V$ normalized to the  
zero-bias resistance $R_0$ measured at low $T=0.2\,{\rm K}\ll T_c$ with  
$R_0= 4.4$ (curve 1), 5.0 (2), 5.3 (3), and $3.6\,\Omega$ (4). The inset shows
a $dV/dI$ curve with $R_0=7.6\,\Omega$ to higher bias.} 
\label{fig1} 
\end{figure} 
Figure 1 shows four representative zero-field spectra 
at low $T\leq 0.2\,{\rm K}\ll T_c$. For comparison, the spectra have
been normalized to the zero-bias resistance $R_0$. Almost 20 contacts were
investigated. 
Typical $R_0$ values for stable contacts range from 0.1 to 40\,$\Omega$.
The main result is that two distinctly different types of spectra
are observed. Curves 1 and 2 represent spectra which exhibit a double-minimum 
structure (type 1), curves 3 and 4 represent spectra which show a single minimum 
in $dV/dI$ centered at $V=0$, and shoulders symmetric in $V$ at higher
bias (type 2). The normal-state background of both types of curves is almost
flat with no additional structure (see inset of Fig.\,1). Minima at finite $V$ 
in the differential resistance are expected for N/S PCs
due to Andreev reflection (AR) at the N/S interface for both 
conventional \cite{blo82} and unconventional superconductors 
\cite{bru90}. This scattering process where an electron is injected
and a hole is retro-reflected with probability $R_A$
leads to a minimum in $dV/dI$ vs $V$ of width $\approx 2\Delta /e$.
A finite but small probability $1-R_A$ of normal 
reflection due to an interface barrier increases the zero-bias 
resistance and leads to the characteristic double-minimum feature. 
Phenomenologically, the barrier strength is modeled by a parameter $Z$ 
which assumes a $\delta$-functional barrier potential at the 
interface. $Z=0$ corresponds to a pure metallic PC without interface
barrier. With increasing $Z$ gradually the tunneling limit is approached. 
Qualitatively, the same general behavior is expected 
for unconventional SC with a $k$-dependent 
gap function $\Delta =\Delta (\bbox{k})$, although the "transparency" of the 
junction for AR has to be determined self-consistently to take into account that the 
interface itself might be pairbreaking for some OP symmetries.

Recently, the conductance spectra of N/S junctions
have been calculated for unitary and non-unitary spin-triplet pairing states
\cite{hon98,yam97}. Taking the quasi-two dimensionality of the system
into account the gap function parametrized by 
a vector function $\bbox{d}(\bbox{k})$ is given by one of the following functions:
\begin{center}
	\begin{tabular}{lc}
	$\bbox{d}(\bbox{k})\sim \bbox{\hat{z}}(k_x + \imath k_y)$\hspace*{5mm}& $(A)$\\
	$\bbox{d}(\bbox{k})\sim \bbox{\hat{x}}k_x + \bbox{\hat{y}}k_y$\hspace*{5mm}& $(B)$\\
	$\bbox{d}(\bbox{k})\sim (\bbox{\hat{x}}+\imath \bbox{\hat{y}})(k_x - 
	\imath k_y)$\hspace*{5mm}& ($C$)\\
	\end{tabular}
\end{center}
Of these three $p$-wave states, $A$ and $B$ are unitary. State $C$ is a
non-unitary state with the consequence that the excitation spectrum is gapped only for one 
spin direction. The pairing states $A$, $B$, and $C$ are realized in the three superfluid 
phases $A$, $B$, and $A_1$ of $^3$He, respectively, 
and are being discussed as possible candidates for the 
OP of Sr$_2$RuO$_4$ \cite{ric95}.
The key result of the calculations \cite{hon98,yam97} is that for current injection 
into the $ab$-plane two types of 
spectra can be obtained depending on $Z$, i.\,e. 
gap-like structures for low-$Z$ contacts as found for many superconductors 
(type 1) and spectra with a ZBA for 
high-$Z$ contacts (type 2). 
The ZBA is due to low-lying ($|\epsilon_b | \ll \Delta)$ Andreev
bound states at the surface caused by a sign change of the pair potential \cite{buc81}. 
Our observation of two distinct types of spectra
strongly supports the existence of an unconventional, probably
spin-triplet, OP in Sr$_2$RuO$_4$.
However, these angle- and spin-averaged curves gave no 
unambiguous criterion to discriminate between the different spin-triplet pairing states,
since for all OP symmetries $A$, $B$, and $C$ the 
theoretically predicted spectra look qualitatively 
the same, while quantitatively the gapless channel of the non-unitary state leads to a 
reduction of the SC-related features.

In order to quantitatively compare the theoretical spectra with our data
and to extract the magnitude and the $T$
dependence of the gap, $\Delta(T)$, one has to perform a more detailed calculation
than the calculations \cite{hon98,yam97} within a Blonder-Tinkham-Klapwijk model 
\cite{blo82}. 
We solved the problem self-consistently for the surface state
using weak-coupling quasiclassical theory \cite{ser83}, which allows conveniently to
self-consistently include effects on the calculated 
spectra of bulk disorder, surface suppression of the OP,
and surface quality.  
Given the smallness of the ratios of $T_c$ to the Fermi temperature, 
$T_c/T_F\sim 10^{-4}$,
and to the paramagnon temperature, $T_c/T_P\sim 10^{-2}$ \cite{maz97}
a weak-coupling theory
should be a good approximation for Sr$_2$RuO$_4$.
A $p$-wave OP 
$\bbox{d}(\bbox{k})\sim \bbox{\hat{z}}(k_x + \imath k_y)$ is assumed. 
The current transport across the
PC is modeled allowing a tunable transparency \cite{zai84}. A phenomenological acceptance cone, 
${\cal D} (\phi) ={\cal D}_o \exp(-\lambda \sin^2 \phi)$, puts emphasis on quasiparticle
transmission through the PC at incidence angles, $\phi$, 
within $\sin^2 \phi \le \lambda^{-1}$ of the contact normal. ${\cal D}_o$ is the
transmission probability for quasiparticles along the contact normal. 
The signal of the measurements
is only 1-5\% of the background conductance, hence the overall amplitude of the
calculated spectra must be rescaled for the comparison to
be made \cite{goll95}.

\begin{figure}[b]
\centerline{\epsfysize=0.5\textwidth \epsfbox{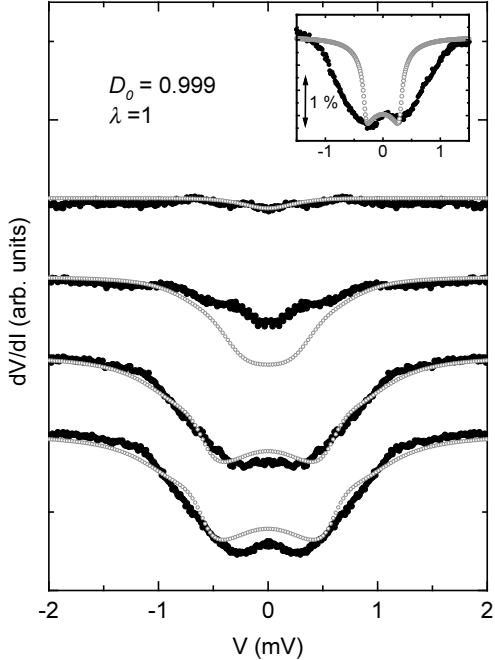}} 
\caption{Temperature dependence of the spectra with a double-minimum structure. 
Closed (open) symbols denote the measured (calculated) spectra. 
The temperatures $T$ (in K, from bottom to top) 
are $T=0.23, 0.40, 1.02$, and $1.25$. For clarity,  
the curves at higher $T$ are shifted with respect to the curve at lowest $T$. 
The inset shows the comparison of the curve at lowest $T$ to a calculation 
within the Blonder-Tinkham-Klapwijk model with an isotropic gap
$\Delta=0.3$\,meV and $Z=0.2$.} 
\label{fig2} 
\end{figure}
\begin{figure} 
\centerline{\epsfysize=0.5\textwidth \epsfbox{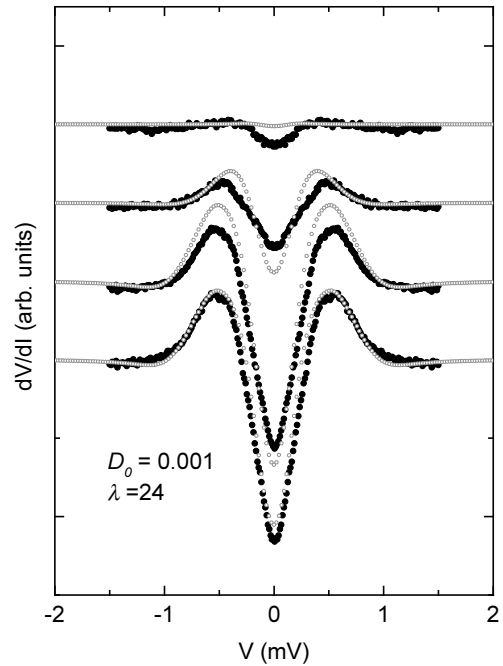}} 
\caption{Temperature dependence of the spectra with a single-minimum.  
Closed (open) symbols denote the measured (calculated) spectra. 
The temperatures $T$ (in K, from bottom to top) 
are $T=0.3, 0.41, 1.01$, and $1.26$. 
For clarity, the curves at higher $T$ are shifted with respect to the curve at lowest $T$.} 
\label{fig3} 
\end{figure}

 
The measured spectra show SC-related structures up to 
$T\approx 1.3$\,K, close to the SC onset
determined resistively. This reduction with respect to
the optimal $T_c$ of 1.5\,K can be attributed to
non-magnetic impurities limiting the quasiparticle mean free path
to 15 coherence lengths ($\xi_o=590$\,\AA) in the superconductor \cite{mac98}.
We first discuss the spectra for $T\ll T_c$.
The existence of two types of spectra is instrumental in determining the 
zero-temperature energy gap, $\Delta(0)$. 
In this respect, it is reassuring for our assignment of $p$-wave
SC that these two different types of spectra yield the same $\Delta_0$
as will be discussed now.
In a simple $s$-wave picture $\Delta(T)$ can be determined
from the position of the minima in $dV/dI$ (see inset of Fig.\,\ref{fig2}). 
Independently of $Z$, the minima occur at $V\approx \pm \Delta /e$. 
However, the spectra of type 1 are much wider than expected for the $s$-wave case,
although the double-minimum feature is reproduced quite well (see inset of Fig.\,\ref{fig2}).
Therefore, we first focus on the low-transmission curves 3 and 4 in Fig.\,\ref{fig1}.
If the quasiparticle transmission is low and transport is 
restricted to small angles of incidence $\phi$,
the conductance spectrum is dominated by the Andreev bound states close 
to the Fermi level \cite{buc81}.
The width and height of the conductance peak depends on the
acceptance cone as the position of the bound states disperses with $\phi$ as
$\epsilon_b(\phi)\approx \Delta \sin \phi$ and the weight 
of the state as $w_b(\phi) \approx |\Delta|\cos\phi$. 
The position of the shoulders where $dV/dI$ levels off towards 
larger $V$, on the other hand, does not depend on details of the acceptance cone. 
It is here where
we can read off directly the magnitude of the energy gap $\Delta_0$. 
The value $2\Delta_0/e=2.2$\,mV together with $\lambda=24$ (reflecting the small
angle of incidence) extracted from the calculations
fit the measured spectra quite well, 
see lowest curve in Fig.\,\ref{fig3}.
Returning to the double-minima
spectra, a fit with a high transmission ($D_0\approx 1$) and a large
acceptance cone $(\lambda =1)$ describes the data quite well, with 
the same value of $2.2$\,mV for $2\Delta_0 /e$ (lowest curve in
Fig.\,\ref{fig2}).  This value, consistently extracted from both types of
spectra, is five times  the value expected from a weak-coupling theory.  
We remark that employing the double-minimum criterion for all spectra of 
type 1 on Sr$_2$RuO$_4$
the average width $2\Delta_0 /e =0.5$\,mV is inconsistent with the large value
obtained. 
In addition, $s$-wave SC cannot
account for the ZBA seen in spectra 3 and 4 in Fig.\,\ref{fig1}.

\begin{figure} 
\centerline{\epsfysize=0.33\textwidth \epsfbox{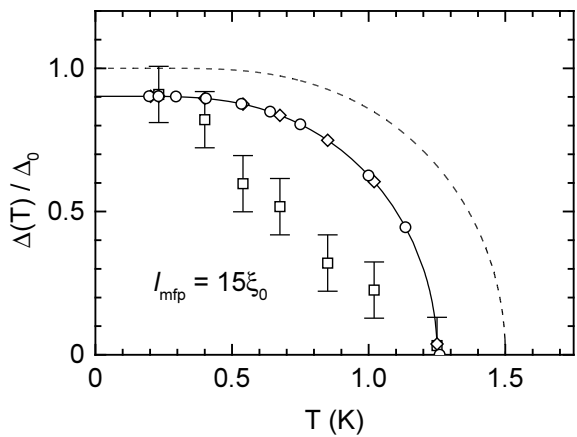}} 
\caption{Temperature dependence of the energy gap $\Delta$ assuming a $p$-wave state for 
curves of type 1 (diamonds) and type 2 (circles) in 
comparison with a $T$ dependence extracted from the excess current (squares). The $T$ 
dependence of a clean $p$-wave gap is indicated by the dashed line, the 
$p$-wave gap modified by impurities by the solid line.} 
\label{fig4} 
\end{figure} 
The $T$ dependence of both types of spectra is shown in 
Figs.\,\ref{fig2} and \ref{fig3} together with calculated spectra.
In both cases, the 
SC-related features become weaker with increasing $T$ and
vanish near $T_c$. $\Delta (T)$ is calculated without
additional parameter 
once $\Delta_0$ has been determined by fitting to the spectra at the lowest $T$. 
The extracted $\Delta (T)$ behaves as the anisotropic gap modified by the 
influence of impurities (solid line), as displayed in Fig.\,\ref{fig4}.
Note that the extracted $\Delta(T)$ is identical for both types of contacts.
We mention that we have chosen the extreme cases $D_0\approx 1$ and 0.001 as 
fitting parameter. Equally good fits are obtained with $D_0\approx 0.9$ and
0.1, respectively.
For a metallic PC, the excess current $I_{exc}$ due to AR
is proportional to $\Delta$ if one assumes an isotropic 
$s$-wave OP \cite{kul92},
and therefore the $T$ dependence of $I_{exc}$ should follow
that of $\Delta$. $I_{exc}(T)$ normalized to the
value $I_{exc}(0)$ at lowest $T$ for the spectra in Fig.\,\ref{fig2} 
vanishes much faster than expected in a weak-coupling theory for an isotropic
gap (Fig.\,\ref{fig4}) and again demonstrates the failure of an
$s$-wave model for the SC in Sr$_2$RuO$_4$.

While we achieve a consistent description within $p$-wave pairing, we note that $d$-wave
pairing as in cuprate superconductors can qualitatively account for a 
ZBA as well \cite{gre99}. $dV/dI$ spectra calculated in a $d_{x^2-y^2}$ 
scenario show a strong dependence
on the relative crystal-to-junction orientation. In particular, junctions
established mainly along the $[100]$ crystal axis should show a largely increased $dV/dI$
at low voltage compared to a sharp minimum along $[110]$. With the 
experimental technique employed by us the current is preferentially injected 
along different crystal axes depending on where the contact is made. 
However, we never observed a large $d_{x^2-y^2}$-derived $dV/dI$ feature associated with the $[100]$
direction. In order to further discriminate between a $d$- and $p$-wave OP,
measurements in a magnetic field are underway.

In summary, directly probing the superconducting energy gap of Sr$_2$RuO$_4$ by means of 
PC spectroscopy gives strong support of an unconventional pairing 
state. The most 
convincing indication clearly comes from the occurrence of a zero-bias 
anomaly for contacts with a weakly-transparent interface, in addition to the more conventional
double-minimum features occurring for highly transparent interface. 
However, even for the latter type of contacts 
the shape of the spectra and their temperature dependence hints at 
unconventional superconductivity. A consistent zero-temperature gap and
$T$ dependence for both types of spectra is obtained within a model of 
$p$-wave pairing.

This work was supported by the Deutsche Forschungsgemeinschaft through
Graduiertenkolleg "Anwend\-un\-gen der Supraleitung".
The work of M.~F. was financed by the EU TMR Network ERBFMRXCT96-0042. 
F.~Li. acknowledges support by the BMBF project No. 13N6918/1.

\end{document}